\documentclass[a4paper,10pt]{article}
\usepackage[centertags]{amsmath}
\usepackage{geometry}
\usepackage{amsfonts}
\usepackage{amssymb}
\usepackage{amsthm}
\usepackage{latexsym}
\usepackage{psfrag}
\renewcommand{\vec}{\mathbf}
\title{Quantum Field Theory Analysis of Polarizations Correlations, Entanglement and Bell's Inequality: Explicit Processes}
\author{N.~Yongram and  E.~B.~Manoukian\footnote{Corresponding author, e-mail: manoukian\_{}eb@hotmail.com}
            \\ The Institute for Fundamental Study, NEP, Naresuan University,\\
           Phitsanulok 65000, Thailand\\
           Fortschr. Phys. \textbf{61}, 668-684 (2013)}
\date{}

\begin{document}

\maketitle
\begin{abstract}
\noindent This paper provides explicit and detailed quantum field theory (QFT) computations of polarizations correlations of emerging particles in several processes in QED, Electroweak Theory, and even in particle productions from strings, and hence are based on dynamical theries. The novel properties observed in the computations of  such polarizations correlations, as predicted by QFT, is that they depend on the speed of the colliding particles, and, in some cases, even on coupling parameters ratios as well as on mass ratios of the underlying theory. These investigations clearly show that arguments based simply on combining spins to generate probabilities of such polarization correlations are not reliable. The novel properties obtained, with details of derivations provided here, hopefully would call for experiments on polarizations correlations which would monitor speeds of colliding particles, and, would, in turn, lead to new tests of the fundamental interactions of QFT such as of QED and of the Electroweak Theory.
\end{abstract}

%\noindent {\bf PACS:} 11.10.Kk, 11.10.-z, 11.25.-w\\

\section{Introduction}
 \noindent Over the years, several experiments have been performed on particles' polarizations correlations (c.f. [1-5]) in the
 light of Bell's inequality (c.f. [6-8]), and other similar experiments have been also proposed in high energy physics (c.f. [9-14]). Quantum field theory, as quantum theory extended to the \textit{relativistic} regime, par excellence, where particles are created and destroyed with the number of particles initially involved not necessarily conserved, is certainly most suitable in providing a theoretical laboratory for investigating such phenomenae by
 carrying \textit{explicit} computations. We have been particularly interested in actual quantum field theory computations of polarization correlations probabilities of particles produced in processes because of the \textit{novelties} we have encountered in such dynamical calculations as opposed to ones obtained na\"{\i}vely by simply combining spins. With well known available and accurate quantum field theory methods of computing processes in high energy physics, methods based on just combining spins, as well as on non-relativistic quantum mechanical methods, would be questionable. Quantum field theory shows that polarization correlations depend, in general, on the \textit{speed} of the particles involved and, in some cases, on the the underlying coupling parameters ratios as well. In all the processes investigated in this report, a clear cut dependence on the speed is encountered, and in one process  dependence not only on energy is obtained, but also on couplings ratios as well as particles mass ratios.\\
 \indent A relevant application of interest of these polarizations correlations probabilities is done in testing Bell's inequality [15,16], by computing, in the process, a
quantity $S$,  referred to as the indicator, which, in a rather \textit{standard} notation, reads
\begin{equation}\label{Eqn1.1}
S=\frac{P_{12}(a_1,a_2)}{P_{12}(\infty,\infty)}-\frac{P_{12}(a_1,a'_2)}{P_{12}(\infty,\infty)}+\frac{P_{12}(a'_1,a_2)}{P_{12}(\infty,\infty)}
+\frac{P_{12}(a'_1,a'_2)}{P_{12}(\infty,\infty)}-\frac{P_{12}(a'_1,\infty)}{P_{12}(\infty,\infty)}-\frac{P_{12}(\infty,a_2)}{P_{12}(\infty,\infty)}.
\end{equation}

\noindent Here $a_1,a_2 (a'_1,a'_2)$ specify directions along which polarizations of two particles, emerging in a process, are measured, which is described here by specific angles, with  $P_{12}(a_1,a_2)/P_{12}(\infty,\infty)$ denoting the joint probability of spins measurements of both particles, and $P_{12}(a_1,\infty)/P_{12}(\infty,\infty)$, $P_{12}(\infty,a_2)/P_{12}(\infty,\infty)$ denoting the probabilities when the polarization of only one of the particles is measured. [Here $P_{12}(\infty,\infty)$ is a normalization factor.] Violation of Bell's inequality means that $S$ falls outside the interval [\,$-$1,0\,] for at least one choice of the variables encountered in these probabilities. And if this happens, such violation provides arguments against so-called local hidden variables theories, in support of the monumental theory we call quantum theory. \\
\indent Exact quantum field theory computations of these probabilities are reported here which have been computed to leading orders in underlying couplings of several processes in QED, Electroweak Theory as well as for some simple particle productions from strings. In all cases violation of Bell's inequality is encountered. It is important to realize, and sometimes not sufficiently emphasized, is that the probabilities in (\ref{Eqn1.1}) are \textit{conditional} probabilities. That is, \textit{given} that the underlying process has occurred, then they denote the probabilities that some of the underlying variables, associated with the particles emerging in the process, have taken on some specific values.\\
\indent The specific processes investigated are best discussed, and spelled out, as we go along through the various sections. For a review of the underlying basic theory of polarization correlations, in the light of Bell's inequality, including the analysis of some elementary processes which, in some very special cases, polarization correlations are independent of energies, see [17]. In the last section, we provide a clear cut conclusion as to what quantum field theory actually says about polarization correlations, in general, which will hopefully open the way to stringent experimental tests.\footnote{Our first explicit \textit{quantum field theory} computations of polarization correlations which show explicit dependence on speed (energy) was done in early 2003 (see Ref. [21]).}\\

\section{$e^+e^- \rightarrow \gamma\gamma$ with initially Polarized or Unpolarized Pair}
\noindent We first consider the process $e^+e^- \rightarrow \gamma\gamma$ [18,19] in the c.m., located at the origin, in QED, with the $e^+, e^-$ initially polarized, with $e^-$ spin up, along the z-axis, and $e^+$ with spin down, and prepared to be moving with momenta,
\begin{equation}\label{Eqn2.1}
\vec{p}_1=\vec{p}(e^-) = \gamma m \beta(0,1,0)=-\vec{p}(e^+)=-\vec{p}_2, \quad \gamma=1/\sqrt{1-\beta^2}.
\end{equation}
\noindent For the spinors we have
\begin{equation}\label{Eqn2.2}
u=\Big(\frac{p^0+m}{2m}\Big)^{1/2}\left(
                                    \begin{array}{c}
                                      \;\left(
                                        \begin{array}{c}
                                          1 \\
                                          0 \\
                                        \end{array}
                                      \right)
                                      \\\\
                                      \text{i} \rho \left(
                                                             \begin{array}{c}
                                                               0 \\
                                                               1 \\
                                                             \end{array}
                                                           \right)
                                       \\
                                    \end{array}
                                  \right),\qquad
                                  v=\Big(\frac{p^0+m}{2m}\Big)^{1/2}\left(
                                    \begin{array}{c}
                                     \text{i} \rho\left(
                                        \begin{array}{c}
                                          0 \\
                                          1 \\
                                        \end{array}
                                      \right)
                                      \\\\
                                      \;\,\left(
                                                             \begin{array}{c}
                                                               1 \\
                                                               0 \\
                                                             \end{array}
                                                           \right)
                                       \\
                                    \end{array}
                                  \right),
                                  \end{equation}
\begin{equation}\label{Eqn2.3}
\rho=\frac{\gamma \beta}{\gamma+1}=\frac{\beta}{1+\sqrt{1-\beta^2}}.
\end{equation}
\indent We consider a process with the momenta of the emerging to be given by
\begin{equation}\label{Eqn2.4}
\vec{k}_1=\gamma m (\sin \theta,0,\cos \theta)=-\vec{k}_2,
\end{equation}
\noindent where we have used the facts that
\begin{equation}\label{Eqn2.5}
|\vec{k}_1|= |\vec{k}_2|=k^0_1=k^0_2=p^0(e^\pm)\equiv p^0=\gamma m.
\end{equation}
\indent The amplitude for this process is given by (c.f.[20])
\begin{equation}\label{Eqn2.6}
M \propto \bar{v}\Big[\frac{\gamma^\mu \gamma k_1 \gamma^\nu}{2 p_1k_1}+\frac{\gamma^\nu \gamma k_2 \gamma^\mu}{2 p_1k_2}+\frac{\gamma^\mu p_1^\nu}{p_1 k_1}+\frac{\gamma^\nu p_1^\mu}{p_1 k_2}\Big]u \,e^\nu_1 e^\mu_2,
\end{equation}
\noindent up to an unimportant numerical factor for the problem at hand. Here $e_1^\nu=(0,\vec{e}_1)$, $e_2^\mu=(0,\vec{e}_2)$ are polarization vectors of the emerging photons
\begin{equation}\label{Eqn2.7}
\vec{e}_j=(-\cos \theta\,\cos\chi_j,\,\sin\chi_j,\,\sin \theta\, \cos \chi_j)\equiv(e^{(1)}_j,e^{(2)}_j,e^{(3)}_j),\quad j=1,2.
\end{equation}
\noindent Given that such a process has occurred, we can ask, rigorously, pertinent questions about the polarization correlations of the photons.\\
\indent To the above end, the \label{Eqn2.8}following matrix elements are readily derived
\begin{align}
\bar{v}(\gamma^i\gamma^0\gamma^j)u&=\frac{p^0+m}{2m}\,2\,\rho\,\text{i} \varepsilon_{ij2},\label{Eqn2.8}\\
\bar{v}\gamma^iu&=\frac{p^0+m}{2m}\,(1-\rho^2)\,\delta^{i3},\label{Eqn2.9}\\
\bar{v}(\gamma^i\gamma^n\gamma^j)u&=\frac{p^0+m}{2m}\,(-\delta^{nj}\delta^{i3}-\delta^{ni}\delta^{j3}+\delta^{ij} \delta^{n3})(1-\rho^2) - \text{i}\frac{p^0+m}{2m}(1+\rho^2)\, \varepsilon_{inj}.\label{Eqn2.10}
\end{align}
\noindent Upon setting,
\begin{equation}\label{Eqn2.11}
\frac{\vec{k}_1}{|\vec{k}_1|}=\vec{n},
\end{equation}
the amplitude $M$ then simplifies to
\begin{equation}\label{Eqn2.12}
M \propto - \text{i} (1+\rho^2) \vec{n}.(\vec{e}_1\times \vec{e}_2) + \beta (1-\rho^2)(e_1^{(2)}e_2^{(3)}+e_1^{(3)}e_2^{(2)}).
\end{equation}
\noindent For $\theta=\pi/2$, this gives
\begin{align}
&\;M \propto - (0,\sin \chi_1,\,\cos \chi_1)_1\,(0,\sin \chi_2,\,\cos\chi_2)_2\nonumber \\
\times\Big\{\text{i}& (1+\rho^2)\Big[\left(
                                      \begin{array}{c}
                                        0 \\
                                        1 \\
                                        0 \\
                                      \end{array}
                                    \right)_1\left(
                                               \begin{array}{c}
                                                 0 \\
                                                 0 \\
                                                 1 \\
                                               \end{array}
                                             \right)_2-\left(
                                                         \begin{array}{c}
                                                           0 \\
                                                           0 \\
                                                           1 \\
                                                         \end{array}
                                                       \right)_1\left(
                                                                  \begin{array}{c}
                                                                    0 \\
                                                                    1 \\
                                                                    0 \\
                                                                  \end{array}
                                                                \right)_2
\Big]\nonumber \\- \beta &(1-\rho^2)\Big[\left(
                                          \begin{array}{c}
                                            0 \\
                                            1 \\
                                            0 \\
                                          \end{array}
                                        \right)_1\left(
                                               \begin{array}{c}
                                                 0 \\
                                                 0 \\
                                                 1 \\
                                               \end{array}
                                             \right)_2+\left(
                                                         \begin{array}{c}
                                                           0 \\
                                                           0 \\
                                                           1 \\
                                                         \end{array}
                                                       \right)_1\left(
                                                                  \begin{array}{c}
                                                                    0 \\
                                                                    1 \\
                                                                    0 \\
                                                                  \end{array}
                                                                \right)_2
\Big]\Big\},\label{Eqn2.13}
\end{align}
\noindent generating a \textit{speed dependent} normalized \textit{entangled} state for the photons given by
\begin{align}
|\phi\rangle = \frac{1}{N}\,\Big\{\text{i}& \frac{(1+\rho^2)}{\sqrt{2}}\Big[\left(
                                      \begin{array}{c}
                                        0 \\
                                        1 \\
                                        0 \\
                                      \end{array}
                                    \right)_1\left(
                                               \begin{array}{c}
                                                 0 \\
                                                 0 \\
                                                 1 \\
                                               \end{array}
                                             \right)_2-\left(
                                                         \begin{array}{c}
                                                           0 \\
                                                           0 \\
                                                           1 \\
                                                         \end{array}
                                                       \right)_1\left(
                                                                  \begin{array}{c}
                                                                    0 \\
                                                                    1 \\
                                                                    0 \\
                                                                  \end{array}
                                                                \right)_2
\Big]\nonumber \\- \beta &\frac{(1-\rho^2)}{\sqrt{2}}\Big[\left(
                                          \begin{array}{c}
                                            0 \\
                                            1 \\
                                            0 \\
                                          \end{array}
                                        \right)_1\left(
                                               \begin{array}{c}
                                                 0 \\
                                                 0 \\
                                                 1 \\
                                               \end{array}
                                             \right)_2+\left(
                                                         \begin{array}{c}
                                                           0 \\
                                                           0 \\
                                                           1 \\
                                                         \end{array}
                                                       \right)_1\left(
                                                                  \begin{array}{c}
                                                                    0 \\
                                                                    1 \\
                                                                    0 \\
                                                                  \end{array}
                                                                \right)_2
\Big]\Big\},\label{Eqn2.14}
\end{align}
\noindent with
\begin{equation}\label{Eqn2.15}
N=[(1+\rho^2)^2 + \beta^2 (1-\rho^2)^2]^{1/2}.
\end{equation}
\indent Therefore, the joint probability of photons polarizations correlations is given by
\begin{equation}\label{Eqn2.16}
P[\chi_1,\chi_2]=\|(0,\,\sin \chi_1, \cos \chi_1)_1\,(0,\,\sin \chi_2, \cos \chi_2)_2 |\phi\rangle\|^2
\end{equation}
\noindent leading to the explicit expression
\begin{equation}\label{Eqn2.17}
P[\chi_1,\chi_2]=\frac{(1+\rho^2)^2\sin^2(\chi_1-\chi_2)+\beta^2 (1-\rho^2)^2 \cos^2(\chi_1+\chi_2)}{2\,[(1+\rho^2)^2+\beta^2(1-\rho^2)^2]}.
\end{equation}
\noindent for all $0 \leq \beta \leq 1$.\\
\indent The four possible angular polarizations directions of the photons are specified by the angles
\begin{equation}\label{Eqn2.18}
(\chi_1,\chi_2),\;\; (\chi_1+\frac{\pi}{2}, \chi_2),\;\; (\chi_1 ,\chi_2+\frac{\pi}{2}),\;\;(\chi_1+\frac{\pi}{2},\chi_2+\frac{\pi}{2}),
\end{equation}
\noindent and it is readily verified that the joint probability $P[\chi_1,\chi_2]$ of polarizations correlations satisfies the important normalization condition
\begin{equation}\label{Eqn2.19}
P[\chi_1,\chi_2]+ P[\chi_1+\frac{\pi}{2}, \chi_2]+ P[\chi_1 ,\chi_2+\frac{\pi}{2}]+P[\chi_1+\frac{\pi}{2},\chi_2+\frac{\pi}{2}]=1.
\end{equation}
\indent If the polarization of only one of the photons, say, photon with associated angle $\chi_1$, is measured, then we have to sum over the two angles $\chi_2,\, \chi_2\,+\,\pi/2$:
\begin{equation}\label{Eqn2.20}
P[\chi_1,-]=P[\chi_1,\chi_2]+P[\chi_1 ,\chi_2+\frac{\pi}{2}],
\end{equation}
\noindent to obtain
\begin{equation}\label{Eqn2.21}
P[\chi_1,-]\,=\,\|(0,\,\sin \chi_1, \cos \chi_1)_1 |\phi\rangle\|^2\,=\,\frac{1}{2},
\end{equation}
\noindent and similarly
\begin{equation}\label{Eqn2.22}
P[-,\chi_2]\,=\,\|(0,\,\sin \chi_2, \cos \chi_2)_2 |\phi\rangle\|^2\,=\,\frac{1}{2}.
\end{equation}
\indent  We note that, in general,
\begin{equation}\label{Eqn12.23}
P[\chi_1,\chi_2] \neq  P[\chi_1,-]P[-,\chi_2],
\end{equation}
\noindent signalling the entanglement of the two photon states, and the non-trivial speed $\beta$ dependence of the probability $P[\chi_1,\chi_2]$ should not be ignored. \textit{Interestingly enough \textit{for}} $\beta\rightarrow0$, \textit{we obtain}
$P(\chi_1,\chi_2) \rightarrow \sin^2(\chi_1-\chi_2)/2$ - \textit{a result which has been known for years}.\\\\
\indent In the standard notation (\ref{Eqn1.1}), we have
\begin{equation}\label{Eqn2.24}
P[\chi_1,\chi_2]=\frac{P_{12}(a_1,a_2)}{P_{12}(\infty,\infty)},\;\;\;P[\chi_1,-]=\frac{P_{12}(a_1,\infty)}{P_{12}(\infty,\infty)},\;\;\;
P[-,\chi_2]=\frac{P_{12}(\infty,a_2)}{P_{12}(\infty,\infty)}.
\end{equation}
\noindent Defining
\begin{equation}\label{Eqn2.25}
S=P[\chi_1,\chi_2]-P[\chi_1,\chi'_2]+P[\chi'_1,\chi_2]+P[\chi'_1,\chi'_2]-P[\chi'_1,-]-P[-,\chi_2],
\end{equation}
\noindent for four angles $\chi_1,\chi_2,\chi'_1,\chi'_2$. The Bell inequality is given by $- 1 \leq S \leq 0$ [15,16]. It is sufficient to realize one experimental situation that violates at least one of the bounds.\\
\indent For all $0 \leq \beta \leq 1$, angles $\chi_1,\chi_2,\chi'_1,\chi'_2$ are readily found leading to a violation of Bell's inequality. For example, for $\beta=0.2$, $\chi_1=0^{\text{o}}$, $\chi_2=23^{\text{o}}$, $\chi'_1=45^{\text{o}}$, $\chi'_2=67^{\text{o}}$, $S=-1.187$ violating the inequality from below.\\

\indent We now consider the process $e^+e^- \rightarrow \gamma\gamma$ [21,19] in the c.m for unpolarized  $e^+, e^-$, which have been prepared to be moving along the z-axis, each with speed $v=\beta$ in opposite directions, prior to their annihilation into a pair of photons, each of momentum $\vec{k}$ in opposite directions. For great generality, we may write for\, $\vec{k}$, and for the photon polarization vectors $\vec{e}_j$, as
\begin{align}
\vec{k}=|\vec{k}|(\cos\phi \sin\theta,\sin\phi \sin\theta,\cos\theta),\, \quad  \vec{e}_j(\lambda_2)=\vec{e}_j(\lambda_1)|_{\chi_j\rightarrow\chi_j+\pi/2},\;\;\;j=1,2,\;\; \quad\text{where},  \label{Eqn2}\\
\vec{e}_j(\lambda_1)=(-\cos\theta \cos\chi_j \cos \phi - \sin\chi_j\sin\phi\,, \sin \chi_j \cos\phi - \cos \theta \cos\chi_j \sin\phi\,,\sin\theta \cos\chi_j).\label{Eqn2.26}
\end{align}
\noindent Here $\chi_j$ are the angles that the polarizations $\vec{e}_j(\lambda_1)$ would make with the x-z plane if $\vec{k}$ where lying in the latter plane. Recall that $\vec{k}.\vec{e}_j(\lambda_i)=0$.\\
\indent In the c.m., for the probability for the process in question, we have (c.f. [22]), up to an unimportant numerical factor for the problem at hand,
\begin{equation}\label{Eqn2.27}
\text{Prob} \propto \frac{1}{4}\frac{(k_1k_2)^2}{(p_1k_1)(p_1k_2)}-\Big(\vec{e}_1.\vec{e}_2+\frac{\vec{e}_1.\vec{p}\,\, \vec{e}_2.\vec{p} \,(k_1k_2)}{(p_1k_1)(p_1k_2)}\Big)^2,
\end{equation}
\noindent by averaging over the initial spin states of the $e^+e^-$ pair. Here $\vec{p}=\vec{p}_1=-\vec{p}_2$, $\vec{k}=\vec{k}_1=-\vec{k}_2$, $p^0=p_1^0=p_2^0=k_1^0=k_2^0$. Upon using the scalar products,
\begin{equation}\label{Eqn2.28}
\vec{e}_1.\vec{p}=|\vec{p}| \sin\theta \cos \chi_1,\quad \vec{e}_2.\vec{p}=|\vec{p}| \sin\theta \cos \chi_2,\quad \vec{p}.\vec{k}_1=|\vec{p}||\vec{k}| \cos \theta=- \vec{p}.\vec{k}_2,
\end{equation}
\noindent making the definitions of the angles $\chi_1,\,\chi_2$ evident, the probability in (\ref{Eqn2}) takes the form
\begin{align}
\text{Prob} \propto \frac{\Big[1-4(1-\beta^2) \cos \chi_1 \cos \chi_2 [\cos(\chi_1-\chi_2)-2 \cos \chi_1 \cos \chi_2]\Big]}{(1-\beta^2 \cos^2 \theta)}\nonumber \\
- \frac{4(1-\beta^2)^2 \cos^2 \chi_1 \cos^2 \chi_2}{(1-\beta^2 \cos^2 \theta)^2}-[\cos(\chi_1-\chi_2)-2 \cos \chi_1 \cos \chi_2]^2, \label{Eqn2.29}
\end{align}
\noindent $\beta=|\vec{p}|/p^0$, $\theta$ is the angle between the vectors $\vec{k}$ and $\vec{p}$.\\
\indent Upon considering the four possible angular polarizations directions of the photons specified by the angles
\begin{equation}\label{Eqn2.30}
(\chi_1,\chi_2),\;\; (\chi_1+\frac{\pi}{2}, \chi_2),\;\; (\chi_1 ,\chi_2+\frac{\pi}{2}),\;\;(\chi_1+\frac{\pi}{2},\chi_2+\frac{\pi}{2}),
\end{equation}
\noindent the following expression for the properly normalized joint probability of the photon polarizations, specified by the angles $\chi_1,\,\chi_2$, emerges
\begin{equation}\label{Eqn2.31}
P[\chi_1,\chi_2]=\frac{1-[\cos(\chi_1-\chi_2)-2 \beta^2 \cos \chi_1 \cos \chi_2]^2}{2[1+2\beta^2(1-\beta^2)]},
\end{equation}
\noindent for all $0 \leq \beta \leq 1$. The proper normalizability of the joint probability distribution
\begin{equation}\label{Eqn2.32}
P[\chi_1,\chi_2]+ P[\chi_1+\frac{\pi}{2}, \chi_2]+ P[\chi_1 ,\chi_2+\frac{\pi}{2}]+P[\chi_1+\frac{\pi}{2},\chi_2+\frac{\pi}{2}]=1,
\end{equation}
\noindent is easily verified.\\
\indent If the polarization of only one of the photons, say, photon with associated angle $\chi_1$, is measured, then we have to sum over the two angles $\chi_2,\, \chi_2\,+\,\pi/2$:
\begin{equation}\label{Eqn2.33}
P[\chi_1,-]=P[\chi_1,\chi_2]+P[\chi_1 ,\chi_2+\frac{\pi}{2}],
\end{equation}
\noindent to obtain
\begin{equation}\label{Eqn2.34}
P[\chi_1,-]=\frac{1+4\beta^2(1-\beta^2)\cos^2\chi_1}{2[1+2\beta^2(1-\beta^2)]},
\end{equation}
\noindent and similarly,
\begin{equation}\label{Eqn2.35}
P[-,\chi_2]=\frac{1+4\beta^2(1-\beta^2)\cos^2\chi_2}{2[1+2\beta^2(1-\beta^2)]},
\end{equation}
\noindent for all $0 \leq \beta \leq 1$. We note that, in general,
\begin{equation}\label{Eqn2.36}
P[\chi_1,\chi_2] \neq  P[\chi_1,-]P[-,\chi_2],
\end{equation}
\noindent signalling the entanglement of the two photon states, and the non-trivial speed $\beta$ dependence of these probabilities should not be ignored. Again  interestingly enough \textit{for} $\beta\rightarrow0$, \textit{we obtain}
 $P(\chi_1,\chi_2) \rightarrow \sin^2(\chi_1-\chi_2)/2$, $P(\chi_1,-)\rightarrow 1/2$, $P(-,\chi_2)\rightarrow 1/2$ - \textit{a result which has been known for years}.\\
\indent In the standard notation (\ref{Eqn1.1}), we have the identifications given in (\ref{Eqn2.24}), with the critical parameter $S$ defined in (\ref{Eqn2.25}).\\
 For example, for $\chi_1=0^\text{o}$, $\chi_2=67^\text{o}$, $\chi'_1=135^\text{o}$, $\chi'_2=23^\text{o}$, give $S=0.207$ for $\beta=0$, violating this inequality from above. For $\chi_1=0^\text{o}$, $\chi_2=23^\text{o}$, $\chi'_1=45^\text{o}$, $\chi'_2=67^\text{o}$, give $S=-1.207$ for $\beta=0$,
violating this inequality from below. Both bounds are violated for these same angles, for all $\beta\leq 0.2$, eliminating local hidden variables theories in support of quantum theory.\\
\indent [\,For completeness we also provide the relevant probabilities for emissions of two photons from spin 0 pairs that we have also carried out:
\begin{align}
P[\chi_1,\chi_2]&=\frac{[\cos(\chi_1-\chi_2)- 2 \beta^2 \cos \chi_1\,\cos \chi_2]^2}{2[1-2 \beta^2(1-\beta^2)]}\label{Eqn2.37}\\
P[\chi_1,-]&=\frac{1-4 \beta^2(1-\beta^2)\cos^2 \chi_1}{2[1-2 \beta^2(1-\beta^2)]}, \label{Eqn2.38}\\
P[-,\chi_2]&=\frac{1-4 \beta^2(1-\beta^2)\cos^2 \chi_2}{2[1-2 \beta^2(1-\beta^2)]}.\label{Eqn2.39}\,]
\end{align}

\section{Elastic $e^-e^-$ Scattering for initially Polarized or Unpolarized Electrons}
\noindent We first consider  the process $e^-e^- \rightarrow e^-e^-$, in the c.m in QED, with initially polarized electrons [18], one with spin up, along the z-axis, and one with spin down. With $\vec{p}_1=\gamma m \beta(0,1,0)=-\vec{p}_2$, denoting the initial momenta of the initial electron and positron respectively, where $\gamma=1/\sqrt{1-\beta^2}$, we consider a process where the momenta of the emerging electrons and are given by
\begin{equation}\label{3.1}
\vec{p}'_1=\gamma m \beta (\sin \theta,0,\cos\theta)=-\vec{p}'_2,
\end{equation}
\noindent where $\theta$ is measured from the z-axis. Given that such a process has occurred, we may ask questions about their final polarization correlations. The spinors of the initial electrons are given by
\begin{equation}\label{Eqn3.2}
u_1=\Big(\frac{p^0+m}{2m}\Big)^{1/2}\left(
                                    \begin{array}{c}
                                      \;\left(
                                        \begin{array}{c}
                                          1 \\
                                          0 \\
                                        \end{array}
                                      \right)
                                      \\\\
                                      \text{i} \rho \left(
                                                             \begin{array}{c}
                                                               0 \\
                                                               1 \\
                                                             \end{array}
                                                           \right)
                                       \\
                                    \end{array}
                                  \right),\qquad
                                 u_2=\Big(\frac{p^0+m}{2m}\Big)^{1/2}\left(
                                    \begin{array}{c}
                                     \;\,\left(
                                        \begin{array}{c}
                                          0 \\
                                          1 \\
                                        \end{array}
                                      \right)
                                      \\\\
                                      \text{i} \rho\left(
                                                             \begin{array}{c}
                                                               1 \\
                                                               0 \\
                                                             \end{array}
                                                           \right)
                                       \\
                                    \end{array}
                                  \right),
                                  \end{equation}
\noindent with $\rho$ defined in (\ref{Eqn2.3}). For the final ones we have
\begin{equation}\label{Eqn3.3}
u'_1=\Big(\frac{p^0+m}{2m}\Big)^{1/2}\left(
                                       \begin{array}{c}
                                        \xi_1 \\
                                        \frac{\vec{\sigma}.\vec{p}'_1}{p^0+m}\, \xi_1\\
                                       \end{array}
                                     \right),\qquad u'_2=\Big(\frac{p^0+m}{2m}\Big)^{1/2}\left(
                                       \begin{array}{c}
                                        \xi_2 \\
                                        -\frac{\vec{\sigma}.\vec{p}'_1}{p^0+m}\, \xi_2\\
                                       \end{array}
                                     \right),
\end{equation}
\noindent where the two-spinors $\xi_1,\,\xi_2$ will be specified later.\\
\indent The amplitude for the process is given by (c.f. [20])
\begin{equation}\label{Eqn3.4}
M \propto \frac{u(p'_1)\gamma^\mu u(p_1)\,\bar{u}(p'_2)\gamma_\mu u(p_2)}{(p'_1-p_1)^2}-\frac{u(p'_2)\gamma^\mu u(p_1)\,\bar{u}(p'_1)\gamma_\mu u(p_2)}{(p'_2-p_1)^2},
\end{equation}
\noindent up to an unimportant numerical factor for the problem at hand.\\
\indent The following matrix elements are needed for further analysis
\begin{align}
\!\!\!\!\!\!\!\!\bar{u}(p'_1)\gamma^0u(p_1)&=\frac{p^0+m}{2m}\,\xi^\dagger_1 \left(
                                                               \begin{array}{c}
                                                                 1+\text{i} \rho^2 \sin \theta \\
                                                                 - \text{i} \rho^2 \cos \theta \\
                                                               \end{array}
                                                          \right),\\
                                                          \bar{u}(p'_2)\gamma^0u(p_2)&=\frac{p^0+m}{2m}\,\xi^\dagger_2 \left(
                                                               \begin{array}{c}
                                                               -\text{i} \rho^2 \cos \theta \\
                                                                 1- \text{i} \rho^2 \sin \theta \\
                                                               \end{array}
                                                          \right),\\
                                                          \bar{u}(p'_1)\gamma^0u(p_2)&=\frac{p^0+m}{2m}\,\xi^\dagger_1 \left(
                                                               \begin{array}{c}
                                                                 \text{i} \rho^2 \cos \theta \\
                                                                 1 + \text{i} \rho^2 \sin \theta \\
                                                               \end{array}
                                                          \right),\\
                                                          \bar{u}(p'_2)\gamma^0u(p_1)&=\frac{p^0+m}{2m}\,\xi^\dagger_2 \left(
                                                               \begin{array}{c}
                                                                 1-\text{i} \rho^2 \sin \theta \\
                                                                  \text{i} \rho^2 \cos \theta \\
                                                               \end{array}
                                                          \right),\\
                                                          \!\!\!\!\bar{u}(p'_1)\gamma^ju(p_1)&=\frac{p^0+m}{2m} \rho\xi^\dagger_1 \Big[\left(
                                                               \begin{array}{c}
                                                                 \text{i}+ \sin \theta \\
                                                                 -  \cos \theta \\
                                                               \end{array}
                                                          \right)\delta^{j1}\!+\!\text{i}\left(
                                                                                       \begin{array}{c}
                                                                                         -\text{i}+ \sin \theta\\
                                                                                          -  \cos \theta \\
                                                                                       \end{array}
                                                                                     \right)\delta^{j2}\!+\!\left(
                                                                                                          \begin{array}{c}
                                                                                                            -\cos \theta \\
                                                                                                            -\text{i}+ \sin \theta\\
                                                                                                          \end{array}
                                                                                                        \right)\delta^{j3}
                                                          \Big],\\
                                                          \!\!\!\!\bar{u}(p'_2)\gamma^ju(p_2)&=\frac{p^0+m}{2m} \rho\xi^\dagger_2 \Big[\left(
                                                               \begin{array}{c}
                                                                -  \cos \theta \\
                                                                 \text{i}-\sin \theta \\
                                                               \end{array}
                                                          \right)\delta^{j1}\!+\!\text{i}\left(
                                                                                       \begin{array}{c}
                                                                                         \cos \theta\\
                                                                                          \text{i}+ \sin \theta\\
                                                                                       \end{array}
                                                                                     \right)\delta^{j2}\!+\!\left(
                                                                                                          \begin{array}{c}
                                                                                                           \text{i}+ \sin \theta \\
                                                                                                            -\cos \theta\\
                                                                                                          \end{array}
                                                                                                        \right)\delta^{j3}
                                                          \Big],\\
                                                           \!\!\!\!\bar{u}(p'_1)\gamma^ju(p_2)&=\frac{p^0+m}{2m} \rho\xi^\dagger_1 \Big[\left(
                                                               \begin{array}{c}
                                                                  \cos \theta \\
                                                                 \text{i}+\sin \theta \\
                                                               \end{array}
                                                          \right)\delta^{j1}\!+\!\text{i}\left(
                                                                                       \begin{array}{c}
                                                                                         -\cos \theta\\
                                                                                          \text{i}- \sin \theta\\
                                                                                       \end{array}
                                                                                     \right)\delta^{j2}\!+\!\left(
                                                                                                          \begin{array}{c}
                                                                                                           \text{i}- \sin \theta \\
                                                                                                            \cos \theta\\
                                                                                                          \end{array}
                                                                                                        \right)\delta^{j3}\Big]\\
                                                                                                         \!\!\!\!\bar{u}(p'_2)\gamma^ju(p_1)&=\frac{p^0+m}{2m} \rho\xi^\dagger_2 \Big[\left(
                                                               \begin{array}{c}
                                                                \text{i}-\sin \theta \\
                                                                 \cos \theta \\
                                                               \end{array}
                                                          \right)\delta^{j1}\!-\!\text{i}\left(
                                                                                       \begin{array}{c}
                                                                                         \text{i}+ \sin \theta\\
                                                                                          - \cos \theta\\
                                                                                       \end{array}
                                                                                     \right)\delta^{j2}\!-\!\left(
                                                                                                          \begin{array}{c}
                                                                                                           \cos \theta \\
                                                                                                         \text{i}+ \sin \theta \\
                                                                                                          \end{array}
                                                                                                        \right)\delta^{j3}\Big].
                                                                                                        \end{align}
\noindent For $\theta=0$, the amplitude $M$ takes the form
\begin{align}
M \propto\xi_1^\dagger\xi_2^\dagger\Big\{(1+6\rho^2+\rho^4)\Big[ \left(\begin{array}{c}
                                                                     0 \\
                                                                     1 \\
                                                                   \end{array}
                                                                 \right)_1\left(
                                                                            \begin{array}{c}
                                                                              1 \\
                                                                              0 \\
                                                                            \end{array}
                                                                          \right)_2-\left(
                                                                   \begin{array}{c}
                                                                     1 \\
                                                                     0 \\
                                                                   \end{array}
                                                                 \right)_1\left(
                                                                            \begin{array}{c}
                                                                              0 \\
                                                                              1 \\
                                                                            \end{array}
                                                                          \right)_2\Big]\nonumber\\
                                                                          +4\, \text{i} \,\rho^2\Big[ \left(\begin{array}{c}
                                                                     0 \\
                                                                     1 \\
                                                                   \end{array}
                                                                 \right)_1\left(
                                                                            \begin{array}{c}
                                                                              0 \\
                                                                              1 \\
                                                                            \end{array}
                                                                          \right)_2+\left(
                                                                   \begin{array}{c}
                                                                     1 \\
                                                                     0 \\
                                                                   \end{array}
                                                                 \right)_1\left(
                                                                            \begin{array}{c}
                                                                              1 \\
                                                                              0 \\
                                                                            \end{array}
                                                                          \right)_2\Big]\Big\},\label{Eqn3.13}
\end{align}
\noindent generating the speed dependent  normalized entangled state of the emerging photons
\begin{align}
|\psi\rangle=\frac{1}{N}\Big\{\frac{(1+6\rho^2+\rho^4)}{\sqrt{2}}\Big[ \left(\begin{array}{c}
                                                                     0 \\
                                                                     1 \\
                                                                   \end{array}
                                                                 \right)_1\left(
                                                                            \begin{array}{c}
                                                                              1 \\
                                                                              0 \\
                                                                            \end{array}
                                                                          \right)_2-\left(
                                                                   \begin{array}{c}
                                                                     1 \\
                                                                     0 \\
                                                                   \end{array}
                                                                 \right)_1\left(
                                                                            \begin{array}{c}
                                                                              0 \\
                                                                              1 \\
                                                                            \end{array}
                                                                          \right)_2\Big]\nonumber\\
                                                                          +\frac{4\, \text{i}\, \rho^2}{\sqrt{2}}\Big[ \left(\begin{array}{c}
                                                                     0 \\
                                                                     1 \\
                                                                   \end{array}
                                                                 \right)_1\left(
                                                                            \begin{array}{c}
                                                                              0 \\
                                                                              1 \\
                                                                            \end{array}
                                                                          \right)_2+\left(
                                                                   \begin{array}{c}
                                                                     1 \\
                                                                     0 \\
                                                                   \end{array}
                                                                 \right)_1\left(
                                                                            \begin{array}{c}
                                                                              1 \\
                                                                              0 \\
                                                                            \end{array}
                                                                          \right)_2\Big]\Big\},\label{Eqn3.14}
\end{align}
\noindent where
\begin{equation}\label{Eqn3.15}
N=[(1+6\rho^2+\rho^4)^2+16 \rho^4]^{1/2},\qquad \xi_j=\frac{1}{\sqrt{2}}\left(
                                                                         \begin{array}{c}
                                                                           \text{e}^{-\text{i}\chi_j/2} \\
                                                                           \text{e}^{+\text{i}\chi_j/2}\\
                                                                         \end{array}
                                                                       \right),\;j=1,2,
                                                                       \end{equation}
\noindent $\rho$ is defined in (\ref{Eqn2.3}), and the angles are measured relative to the x-axis in planes parallel to the x-y one..\\
\indent The joint probability of the polarization correlations of the emerging electrons is then given by
\begin{equation}\label{Eqn3.16}
 P[\chi_1,\chi_2]=\|\xi_1^\dagger\xi_2^\dagger |\psi\rangle\|^2= \frac{\Big[ (1+6\rho^2+\rho^4)\sin\Big(\frac{\chi_1-\chi_2}{2}\Big)- 4 \rho^2 \cos\Big(\frac{\chi_1+\chi_2}{2}\Big)\Big]^2}{2\,[(1+6\rho^2+\rho^4)^2+16 \rho^4\Big]}.
 \end{equation}
\noindent It satisfies the normalization condition
\begin{equation}\label{Eqn3.17}
P[\chi_1,\chi_2]+ P[\chi_1+\pi, \chi_2]+ P[\chi_1 ,\chi_2+\pi]+P[\chi_1+\pi,\chi_2+\pi]=1.
\end{equation}
\indent If only one of the spins polarizations is measured, then we have
\begin{align}
P[\chi_1,-]&=\|\xi_1^\dagger|\psi\rangle\|^2\equiv P[\chi_1,\chi_2]+ P[\chi_1, \chi_2+\pi]\nonumber\\
&=\frac{1}{2}-\frac{4 \rho^2(1+6\rho^2+\rho^4)}{(1+6\rho^2+\rho^4)^2+16 \rho^4}\,\sin \chi_1,\label{Eqn3.18}
\end{align}
\noindent and similarly
\begin{align}
P[-,\chi_2]=\frac{1}{2}+\frac{4 \rho^2(1+6\rho^2+\rho^4)}{(1+6\rho^2+\rho^4)^2+16 \rho^4}\,\sin \chi_2.\label{Eqn3.19}
\end{align}
\indent \noindent For all $0 \leq \beta \leq 1$, angles $\chi_1,\chi_2,\chi'_1,\chi'_2$ are readily found leading to a violation of Bell's inequality. For example, for $\beta=0.3$, $\chi_1=0^{\text{o}}$, $\chi_2=137^{\text{o}}$, $\chi'_1=12^{\text{o}}$, $\chi'_2=45^{\text{o}}$, $S=-1.79$ violating the inequality from below.\\

\indent For initially unpolarized electrons [18], with momenta $\vec{p}_1=\gamma m \beta (0,1,0)= - \vec{p}_2$, we consider a process with electrons
with final momenta
\begin{equation}\label{Eqn3.21}
\vec{p}'_1=\gamma m \beta (1,0,0)= - \vec{p}'_2.
\end{equation}
\noindent Given that such a process has occurred, we enquire about the polarization correlations of the emerging electrons.\\
\indent To the above end, the four spinors of the emerging electrons may be written as
\begin{align}
&u(p'_1)=\Big(\frac{p^0+m}{2m}\Big)^{1/2}\left(
                                       \begin{array}{c}
                                        \xi_1 \\
                                        \frac{\vec{\sigma}.\vec{p}'_1}{p^0+m}\, \xi_1\\
                                       \end{array}
                                     \right),\quad \xi_1=\left(
                                                           \begin{array}{c}
                                                             - \text{i} \cos(\chi_1/2) \\
                                                             \sin(\chi_1/2) \\
                                                           \end{array}
                                                         \right)\\
                                     &u(p'_2)=\Big(\frac{p^0+m}{2m}\Big)^{1/2}\left(
                                       \begin{array}{c}
                                        \xi_2, \\
                                        -\frac{\vec{\sigma}.\vec{p}'_1}{p^0+m}\, \xi_2\\
                                       \end{array}
                                     \right),\quad \xi_2=\left(
                                                           \begin{array}{c}
                                                             - \text{i} \cos(\chi_2/2) \\
                                                             \sin(\chi_2/2) \\
                                                           \end{array}
                                                         \right).
                                                         \end{align}

\indent A tedious computation of the probability of occurrence of the process leads to
\begin{equation}\label{Eqn3.23}
\text{Prob} \propto (1-\beta^2)(1+3\beta^2) \sin^2\Big(\frac{\chi_1-\chi_2}{2}\Big)+\beta^4 \cos^2\Big(\frac{\chi_1+\chi_2}{2}\Big)+ 4 \beta^4\equiv F[\chi_1,\chi_2].
\end{equation}
\indent Given that the process has occurred, the conditional probability that the spin of the emerging electrons make angles $\chi_1,\,\chi_2$ with the z-axis, is then given by
\begin{equation}\label{Eqn3.24}
P[\chi_1,\chi_2]=\frac{F[\chi_1,\chi_2]}{N(\beta)},
\end{equation}
\noindent where $N(\beta)$ is the normalization factor
\begin{align}
\!\!N(\beta) \!=\! F[\chi_1,\chi_2]+F[\chi_1+\pi,\chi_2]+F[\chi_1,\chi_2+\pi]+F[\chi_1+\pi,\chi_2+\pi] = 2 (1+2 \beta^2 + 6 \beta^4),\label{Eqn3.25}
\end{align}
\noindent giving
\begin{equation}\label{Eqn3.26}
P[\chi_1,\chi_2]=\frac{(1-\beta^2)(1+3\beta^2) \sin^2\Big(\frac{\chi_1-\chi_2}{2}\Big)+\beta^4 \cos^2\Big(\frac{\chi_1+\chi_2}{2}\Big)+ 4 \beta^4}{2 (1+2 \beta^2 + 6 \beta^4)}.
\end{equation}
\indent For the probabilities of a measurement of a single polarization, we have
\begin{align}
P[\chi_1,-]=P[\chi_1,\chi_2]+P[\chi_1,\chi_2+\pi]\,=\,\frac{1}{2}, \label{Eqn3.27}\\
P[-,\chi_2]=P[\chi_1,\chi_2]+P[\chi_1+\pi,\chi_2]\,=\,\frac{1}{2}. \label{Eqn3.28}
\end{align}
\indent Using our earlier notation, it is easily verified that Bell's inequality is violated, for example, for $\beta=0.3$, $\chi_1=0^{\text{o}}$, $\chi_2=45^{\text{o}}$, $\chi'_1=90^{\text{o}}$, $\chi'_2=135^{\text{o}}$, giving $S=-1.165$.\\

\section{Elastic $e^+e^-$ Scattering for initially Polarized and Unpolarized Pairs}
\noindent We first consider $e^+e^-$ elastic scattering in QED, with an initially polarized pair [23] with the electron spin up along the z-axis, and the positron  spin down. The electron and positron are prepared to have momenta
$\vec{p}_1=\gamma m \beta (0,1,0)=-\vec{p}_2$, respectively. Consider the process with an emerging pair having momenta $\vec{k}_1=\gamma m \beta (0,0,1)=-\vec{k}_2$. For the two-spinors of the emerging electron and positron, we choose
\begin{equation}\label{Eqn4.1}
\xi_j=\frac{1}{\sqrt{2}}\left(\begin{array}{c}\text{e}^{-\text{i}\chi_j/2} \\
                                                                           \text{e}^{+\text{i}\chi_j/2}\\
                                                                         \end{array}
                                                                       \right),\qquad j=1,2,
                                                                       \end{equation}
\noindent where the angles $\chi_1,\,\chi_2$ are measured from the x-axis in planes parallel to the x-y one. A tedious computation for the probability of occurrence of the process gives
\begin{align}
M \propto &\Big[A(\beta) \cos\Big(\frac{\chi_1+\chi_2}{2}\Big)+B(\beta) \sin\Big(\frac{\chi_1-\chi_2}{2}\Big)\Big]\nonumber\\
    +\,\text{i}\,&\Big[C(\beta) \sin\Big(\frac{\chi_1+\chi_2}{2}\Big)+D(\beta) \cos\Big(\frac{\chi_1-\chi_2}{2}\Big)\Big],\label{Eqn4.2}
\end{align}
\noindent where
\begin{align}
A(\beta)&=1-\rho^2(1-\rho) + 2 \beta^2(1-\rho^2)^2, \label{Eqn4.3}\\
B(\beta)&=\rho(1+\rho) + 8 \beta^2 \rho^2, \label{Eqn4.4}\\
C(\beta)&=1+\rho^2(1-\rho) + 2 \beta(1-\rho^4), \label{Eqn4.5}\\
D(\beta)&=\rho(1+\rho). \label{Eqn4.6}\\
\end{align}
\noindent and $\rho$ is defined in (\ref{Eqn2.3}).\\

\indent Using the notation $F[\chi_1,\chi_2]$ for the absolute value square of the expression on the right-hand side of (69), the joint probability of measurements of the spins of the emerging pair, specified by the angles $\chi_1,\,\chi_2$, is then given by
\begin{equation}\label{Eqn4.7}
P[\chi_1,\chi_2]=\frac{F[\chi_1,\chi_2]}{N(\beta)},
\end{equation}
\noindent where
\begin{align}
N(\beta)&=F[\chi_1,\chi_2]+F[\chi_1+\pi,\chi_2]+F[\chi_1,\chi_2+\pi]+F[\chi_1+\pi,\chi_2+\pi]\nonumber \\
&=2[A^2(\beta)+B^2(\beta)+C^2(\beta)+D^2(\beta)].\label{Eqn4.8}
\end{align}
\noindent Thus we obtain
\begin{align}
P[\chi_1,\chi_2]=\frac{\Big[A(\beta) \cos\Big(\frac{\chi_1+\chi_2}{2}\Big)+B(\beta) \sin\Big(\frac{\chi_1-\chi_2}{2}\Big)\Big]^2}
{2[A^2(\beta)+B^2(\beta)+C^2(\beta)+D^2(\beta)]}\nonumber \\
+ \frac{\Big[C(\beta) \sin\Big(\frac{\chi_1+\chi_2}{2}\Big)+D(\beta) \cos\Big(\frac{\chi_1-\chi_2}{2}\Big)\Big]^2}
{2[A^2(\beta)+B^2(\beta)+C^2(\beta)+D^2(\beta)]}. \label{Eqn4.9}
\end{align}
\indent If only one of the spins is measured, we then have
\begin{align}
P[\chi_1,-]=\frac{1}{2}+ \frac{[A(\beta)B(\beta)+C(\beta)D(\beta)]\,\sin \chi_1}{[A^2(\beta)+B^2(\beta)+C^2(\beta)+D^2(\beta)]},\label{Eqn4.10}\\
P[-,\chi_2]=\frac{1}{2}- \frac{[A(\beta)B(\beta)-C(\beta)D(\beta)]\,\sin \chi_2}{[A^2(\beta)+B^2(\beta)+C^2(\beta)+D^2(\beta)]}.\label{Eqn4.11}
\end{align}
\indent For all $0 \leq \beta \leq 1$, angles $\chi_1,\chi_2,\chi'_1,\chi'_2$ are readily found leading to a violation of Bell's inequality.
For example, for $\beta=0.9$, $\chi_1=0^{\text{o}}$, $\chi_2=45^{\text{o}}$, $\chi'_1=69^{\text{o}}$, $\chi'_2=200^{\text{o}}$, $S=-1.311$.\\
\indent We now consider initially unpolarized $e^+e^-$ pair. For the final pair, their corresponding spinors are set to be

\begin{align}
&u(k_1)=\Big(\frac{k^0+m}{2m}\Big)^{1/2}\left(
                                       \begin{array}{c}
                                        \xi_1 \\
                                        \frac{\vec{\sigma}.\vec{k}_1}{k^0+m}\, \xi_1\\
                                       \end{array}
                                     \right),\quad \xi_1=\left(
                                                           \begin{array}{c}
                                                             - \text{i} \cos(\chi_1/2) \\
                                                             \sin(\chi_1/2) \\
                                                           \end{array}
                                                         \right)\\
                                     &v(k_2)=\Big(\frac{k^0+m}{2m}\Big)^{1/2}\left(
                                       \begin{array}{c}
                                        -\frac{\vec{\sigma}.\vec{k}_1}{k^0+m}\, \xi_2\\
                                       \;\, \xi_2\\
                                       \end{array}
                                     \right),\quad \xi_2=\left(
                                                           \begin{array}{c}
                                                             - \text{i} \cos(\chi_2/2) \\
                                                             \sin(\chi_2/2) \\
                                                           \end{array}
                                                         \right).
                                                         \end{align}
\indent The probability of the occurrence of the process is worked out to be
\begin{align}
&\text{Prob} \propto [2\,\beta^4(1+2\beta^2)-3(1+\beta^2)]\,\sin^2\Big(\frac{\chi_1-\chi_2}{2}\Big)\nonumber\\
+ (1&+\beta^2 +2 \beta^4)\,\cos^2\Big(\frac{\chi_1+\chi_2}{2}\Big)+5(1-\beta^2)\equiv F[\chi_1,\chi_2]. \label{Eqn4.13}
\end{align}
\noindent and
\begin{align}
F[\chi_1,\chi_2]+F[\chi_1+\pi,\chi_2]+F[\chi_1,\chi_2+\pi]+F[\chi_1+\pi,\chi_2+\pi]\nonumber \\
=8[2-3\beta^2+\beta^4+\beta^6]\equiv N(\beta). \label{Eqn4.14}
\end{align}
\indent The joint probability $P[\chi_1,\chi_2]$ is then given by
\begin{align}
P[\chi_1,\chi_2]=\frac{[2\,\beta^4(1+2\beta^2)-3(1+\beta^2)]}{N(\beta)}\,\sin^2\Big(\frac{\chi_1-\chi_2}{2}\Big)\nonumber\\
+\frac{[1+\beta^2 +2 \beta^4]}{N(\beta)}\,\cos^2\Big(\frac{\chi_1+\chi_2}{2}\Big)+\frac{5(1-\beta^2)}{N(\beta)}.\label{Eqn4.15}
\end{align}
\indent For the probabilities corresponding to the measurements of only one of the spins, we simply have
\begin{equation}\label{Eqn4.16}
P[\chi_1,-]=\frac{1}{2},\qquad P[-\chi_2]=\frac{1}{2}.
\end{equation}
\indent For example, for $\beta=0.8$, $\chi_1=0^{\text{o}}$, $\chi_2=45^{\text{o}}$, $\chi'_1=210^{\text{o}}$, $\chi'_2=15^{\text{o}}$, $S=-1.167$.\\

\section{Muon Pair Production in the Standard Electroweak Theory}
\noindent Now we come to a very interesting process which will turn up to be quite important in the present investigations. We consider the process $e^-e^+ \rightarrow \mu^- \mu^+$ in the Standard Electroweak theory [24]. At the outset we notice here that since the mass of the electron is much smaller than the one of the muon, sufficient energy is needed to create the muons. That is, the energy of the $e^-e^+$ is critical for the occurrence of such a process. Is this dependence reflected in the polarization correlations of the muons?. It is indeed, as we will see below. In the c.m., we choose the momenta of the electron and positron to be, respectively, $\vec{p}=\gamma m_e \beta (0,1,0)=-\vec{k}$, with $m_e$, here,  denoting the mass of the electron. The momenta of the emerging $\mu^-, \mu^+$ will be taken to be
\begin{equation}\label{Eqn5.1}
\vec{p}'=\gamma'm_\mu \beta'(1,0,0)=-\vec{k}',
\end{equation}
\noindent where $m_\mu$ is the mass of the muon. With the spins of $e^-$, $e^+$, prepared with respect to the z-axis, we write for their corresponding spinors
\begin{equation}\label{Eqn5.2}
u(p)=\sqrt{\frac{\gamma+1}{2}}\left(
                                \begin{array}{c}
                                  \qquad\left(
                                     \begin{array}{c}
                                       1 \\
                                       0 \\
                                     \end{array}
                                   \right)
                                   \\\\
                                  \text{i} \frac{\gamma \beta}{\gamma+1}\left(
                                                                          \begin{array}{c}
                                                                            0 \\
                                                                            1 \\
                                                                          \end{array}
                                                                        \right)
                                   \\
                                \end{array}
                              \right),\quad v(k)=\sqrt{\frac{\gamma+1}{2}}\left(
                                \begin{array}{c}
                                  \text{i} \frac{\gamma \beta}{\gamma+1}\left(
                                     \begin{array}{c}
                                       1 \\
                                       0 \\
                                     \end{array}
                                   \right)
                                   \\\\
                                  \qquad\left(
                                                                          \begin{array}{c}
                                                                            0 \\
                                                                            1 \\
                                                                          \end{array}
                                                                        \right)
                                   \\
                                \end{array}
                              \right).
                              \end{equation}
\noindent Obviously, there is a non-zero probability of occurrence of the above process. Given that such a process has occurred, we compute the conditional joint probability of spins measurement of $\mu^-$, $\mu^+$ along directions specified by angles $\chi_1$, $\chi_2$ measured from the z-axis. Here we have considered the singlet state.\\
\indent A fairly tedious computation for the invariant amplitude of the process [23,24], leads for the amplitude of occurrence, which when written in terms of the energy $\zeta$ available initially in the system, takes the form
\begin{align}
 M \propto &\Big[A(\zeta) \sin\Big(\frac{\chi_1-\chi_2}{2}\Big) + B(\zeta)\sin\Big(\frac{\chi_1+\chi_2}{2}\Big)+C(\zeta) \cos\Big(\frac{\chi_1-\chi_2}{2}\Big)\Big]\nonumber \\
 -\, \text{i} &\Big[D(\zeta)\sin\Big(\frac{\chi_1+\chi_2}{2}\Big)+E(\zeta) \cos\Big(\frac{\chi_1-\chi_2}{2}\Big)\Big],\label{Eqn5.3}
                               \end{align}
\noindent where
\begin{align}
A(\zeta)&=\Big(\frac{M^2_Z}{4 \zeta^2} + a b^2 - 1\Big),\label{Eqn5.4}\\
B(\zeta)&=-\Big(\frac{m_e}{m_\mu}\Big)\,\Big(\frac{M^2_Z}{4 \zeta^2} + a b^2 - 1\Big),\label{Eqn5.5}\\
C(\zeta)&=\frac{a b \,m_e}{\zeta m_\mu}\sqrt{\zeta^2-m_\mu^2}, \label{Eqn5.6}\\
D(\zeta)&=\frac{a}{m_\mu \zeta}\sqrt{\zeta^2-m_\mu^2}\,\sqrt{\zeta^2-m_e^2}, \label{Eqn5.7}\\
E(\zeta)&=-\frac{ab}{m_\mu}\sqrt{\zeta^2-m_e^2}, \label{Eqn5.8}
\end{align}
\noindent and
\begin{equation}\label{Eqn5.9}
a\equiv \frac{g^2}{16\,e^2 \cos^2 \theta_W}\cong0.353,\qquad b\equiv 1-4 \sin^2 \theta_W \cong 0.08,
\end{equation}
\noindent where the symbols in (\ref{Eqn5.9}) have their usual meanings.\\
\indent Denoting the absolute value squared of the right-hand side of (\ref{Eqn5.3}) by  $F[\chi_1,\chi_2]$, we obtain
\begin{align}
F[\chi_1,\chi_2]+F[\chi_1+\pi,\chi_2]+F[\chi_1,\chi_2+\pi]+F[\chi_1+\pi,\chi_2+\pi]\nonumber \\
=2\,[A^2(\zeta)+B^2(\zeta)+C^2(\zeta)+D^2(\zeta)+E^2(\zeta)]\equiv N(\zeta). \label{Eqn5.10}
\end{align}
\indent The relevant probabilities are then
\begin{align}
P[\chi_1,\chi_2]=\frac{1}{N(\zeta)}\Big[A(\zeta)\sin\Big(\frac{\chi_1-\chi_2}{2}\Big)+B(\zeta)\sin\Big(\frac{\chi_1+\chi_2}{2}\Big)+
C(\zeta)\cos\Big(\frac{\chi_1-\chi_2}{2}\Big)\Big]^2\nonumber \\
+ \frac{1}{N(\zeta)}\Big[D(\zeta)\sin\Big(\frac{\chi_1+\chi_2}{2}\Big)+
E(\zeta)\cos\Big(\frac{\chi_1-\chi_2}{2}\Big)\Big]^2, \label{Eqn5.11}
\end{align}
\begin{equation}\label{Eqn5.12}
P[\chi_1,-]=\frac{1}{2}-\frac{2\,B(\zeta)}{N(\zeta)}\,[A(\zeta)\,\cos \chi_1+C(\zeta) \sin \chi_1],
\end{equation}
\begin{equation}\label{Eqn5.13}
P[-,\chi_2]=\frac{1}{2}+\frac{2\,B(\zeta)}{N(\zeta)}\,[A(\zeta)\,\cos \chi_2+C(\zeta) \sin \chi_2],
\end{equation}
\indent It is important to note that
\begin{equation}\label{Eqn5.14}
P[\chi_1,\chi_2]\,\neq\,P[\chi_1,-]\,P[-,\chi_2],
\end{equation}
\noindent in general, showing an obvious correlation occurring between the two spins.\\
\indent The indicator $S$ is readily computed from the above probabilities since the explicit expressions for $A(\zeta),\,B(\zeta),\,C(\zeta),\,
D(\zeta),\,E(\zeta)$ are known. For $\zeta=105.656$ MeV, i.e., near threshold, an optimal value of $S$ is obtained to be equal to $-$ 1.28203, for
 $\chi_1=0^{\text{o}}$, $\chi_2=45^{\text{o}}$, $\chi'_1=90^{\text{o}}$, $\chi'_2=135^{\text{o}}$, clearly violating Bell's inequality. For energies originally carried out in the experiment of the differential cross section of the process in question, $\zeta\sim$ 34 GeV, an optimal value of $S$ is obtained to be equal to $-$ 1.22094, for $\chi_1=0^{\text{o}}$, $\chi_2=45^{\text{o}}$, $\chi'_1=51.13^{\text{o}}$, $\chi'_2=170.85^{\text{o}}$, leading again to a violation of Bell's inequality.\\
 \indent It is well known [27] that the process $e^-e^+ \rightarrow \mu^-\mu^+$ as computed in the Electroweak theory is, in general, in much better agreement with experiment than that of a QED computation. Near threshold, the indicator $S_{\text{QED}}$, however, computed within QED coincides with that of the indicator $S$ obtained above in the Electroweak theory, and varies slightly at higher energies.\\

 \section{Pair Productions From Strings}
 \noindent For completeness, we finally consider polarization correlations of pair productions [28] via classical strings. We investigate the polarization correlations of $e^-e^+$ pair productions via charged and neutral Nambu strings via processes of photon and graviton emissions, respectively. We consider circularly oscillating closed strings as a cylindrical symmetric solution [29-35] arising from the Nambu action [32-39], as perhaps the simplest string structure for field theory studies. Explicit expressions for the joint probabilities as well as of the probabilities corresponding to a measurement of only one of the spins are derived and again are found to be speed dependent. Due to the different expressions of the probabilities obtained, in general, for charged and neutral strings, inquiries about such correlations, would indicate whether the string is charged or uncharged.\\
 \indent The trajectory of the closed string is described by a vector function $\vec{R}(\sigma,t)$, where $\sigma$ parametrizes the string, satisfying the relations
 \begin{align}
\ddot{\vec{R}}-\vec{R}''=&\,\vec{0}, \quad \dot{\vec{R}}.\vec{R}'=0, \quad \dot{\vec{R}}^2+\vec{R}'^2=1,\label{Eqn6.1}\\
&\vec{R}\Big(\sigma+\frac{2\pi}{m},t\Big)=\vec{R}(\sigma,t), \label{Eqn6.2}
 \end{align}
\noindent where the mass scale $m$ is taken to be the mass of the electron, $\dot{\vec{R}}$ denotes the time derivative $\partial \vec{R}/\partial t$, $\vec{R}'=\partial \vec{R}/\partial \sigma$, with the general solution
\begin{equation}\label{Eqn6.3}
\vec{R}(\sigma,t)=\frac{1}{2}\,[\vec{A}(\sigma-t)+\vec{B}(\sigma+t)], \quad\vec{A}'^2+\vec{B}'^1=1.
\end{equation}
 \indent We consider a solution of the form [30-35]
\begin{equation}\label{Eqn6.4}
\vec{R}(\sigma,t)=\frac{1}{m}\,[\,\cos m \sigma,\,\sin m \sigma,0\,]\,\sin m t,
\end{equation}
\noindent with the z-axis perpendicular to the plane of oscillations.\\
\indent For a string of total charge $Q$, this generates a current density [33] $J^\mu(x)$ with structure ($x=(t,\vec{r},z)$)
\begin{align}
J^\mu(x)&= \int \frac{\text{d}^2\vec{p}}{(2\pi)^2}\int_{-\infty}^\infty\frac{\text{d}q}{2\pi}\int_{-\infty}^\infty \frac{\text{d}P^0}{2\pi}\,
\text{e}^{\text{i} \vec{p}.\vec{r}}\text{e}^{\text{i} qz}\text{e}^{-\text{i} P^0t}\,J^\mu(P^0,\vec{p}),\label{Eqn6.5}\\
J^\mu(P^0,\vec{p})&=2\,\pi\sum_N\delta(P^0-m N)\,B^\mu(\vec{p},N),\label{Eqn6.6}\\
B^0(\vec{p},N)&=a_N\,J^2_{N/2}\Big(\frac{|\vec{p}|}{2m}\Big),\qquad a_N=Q\,(-1)^{N/2}\cos \Big(\frac{N \pi}{2}\Big),\label{Eqn6.7}\\
\vec{B}(\vec{p},N)&=\frac{m N}{|\vec{p}|^2}\,\vec{p}\,B^0(\vec{p},N),\label{Eqn6.8}
\end{align}
\noindent where the summation in (\ref{Eqn6.6}), is over all integers $N$, and the $J_{N/2}$ are the ordinary Bessel functions of order $N/2$.\\
\indent We consider $e^-,e^+$ pair production from a photon emitted from the charged string. The corresponding amplitude [40,41] is then given y
\begin{equation}\label{Eqn6.9}
M \propto J^\mu(2p^0,\vec{p}_1+\vec{p}_2)\,\frac{1}{(p_1+p_2)^2}\,[\bar{u}(p_1,\sigma_1)\gamma_\mu u(p_2,\sigma_2)],
\end{equation}
\noindent with the four momenta of $e^-,e^+$, respectively, given by
\begin{align}
&\vec{p}_1=k(0,1,0),\qquad \;\vec{p}_2=k(1,0,0),\qquad \; k=m \gamma \beta, \label{Eqn6.10}\\
&p_1^0=p_2^0=\sqrt{k^2+m^2}\equiv p^0=m \gamma,\quad \gamma=1/\sqrt{1-\beta^2}. \label{Eqn6.11}
\end{align}
The measurement of the spin projection of the electron is taken along an axis making an angle $\chi_1$ with the z-axis and lying in a plane parallel to the x-z plane
\begin{equation}\label{Eqn6.12}
u=\Big(\frac{p^0+m}{2m}\Big)^{1/2}
             \left(
                \begin{array}{c}
                  \xi_1 \\
                  \frac{k\,\sigma_2}{p^0+m}\,\xi_1 \\
                \end{array}
              \right),\qquad v=\Big(\frac{p^0+m}{2m}\Big)^{1/2}
             \left(
                \begin{array}{c}
                  -\frac{k\,\sigma_2}{p^0+m}\,\xi_2 \\
                  \xi_2 \\
                \end{array}
              \right),
              \end{equation}
where the direction of spin projection of the positron lies in a plane parallel to the y-z plane. For the 2-spinors, we have
\begin{equation}\label{Eqn6.13}
\xi_1=\left(
        \begin{array}{c}
          \cos \Big(\frac{\chi_1}{2}\Big) \\
          - \sin \Big(\frac{\chi_1}{2}\Big)\\
        \end{array}
      \right),\qquad \xi_2=\left(
                            \begin{array}{c}
                              \sin \Big(\frac{\chi_2}{2}\Big) \\
                              \cos \Big(\frac{\chi_2}{2}\Big)\\
                            \end{array}
                          \right).
                          \end{equation}
\indent A tedious computation yields
\begin{align}
M \propto \frac{\Big[- \text{i} \Big(1-\frac{\gamma \beta^2}{2}\Big)\cos \Big(\frac{\chi_1-\chi_2}{2}\Big)
+\Big(1+\frac{\gamma \beta^2}{2}\Big)\cos \Big(\frac{\chi_1+\chi_2}{2}\Big)\Big]}{(\gamma^2\beta^2-2)}\sum_N\delta(2p^0-m N)B^0, \label{Eqn6.14}
\end{align}
\noindent where note that $2p^0/m$ is quantized.\\
\indent Given that the above process has occurred, the joint probability of measurements of the spins of the pair becomes
\begin{equation}\label{Eqn6.15}
P[\chi_1,\chi_2]=\frac{\big(2\sqrt{1-\beta^2}-\beta^2\big)^2\cos^2 \Big(\frac{\chi_1-\chi_2}{2}\Big)
+ \big(2\sqrt{1-\beta^2}+\beta^2\big)^2\cos^2 \Big(\frac{\chi_1+\chi_2}{2}\Big)}{4(2-\beta^2)^2}.
\end{equation}
\noindent The probabilities associated with the measurement of only one of the spins become simply
\begin{equation}\label{Eqn6.16}
P[\chi_1,-]\,=\,\frac{1}{2},\qquad P[-,\chi_2]\,=\,\frac{1}{2}.
\end{equation}
\indent For example, for for $\beta=0.8$, $\chi_1=0^{\text{o}}$, $\chi_2=160^{\text{o}}$, $\chi'_1=100^{\text{o}}$, $\chi'_2=10^{\text{o}}$, give for the indicator $S=-$ 1.088, violating Bell's inequality.\\
\indent It is interesting to note that in the extreme relativistic case $\beta \rightarrow 1$
\begin{equation}\label{Eqn6.17}
P[\chi_1,\chi_2]=\frac{1}{4}\Big[\cos^2 \Big(\frac{\chi_1-\chi_2}{2}\Big)
+\cos^2 \Big(\frac{\chi_1+\chi_2}{2}\Big)\Big].
\end{equation}
\noindent This will be compared with the one resulting from a neutral string.\\
\indent A neutral string of mass $M$, generates an energy-momentum tensor density $T^{\mu\nu}$ with structure [34]
\begin{align}
T^{\mu\nu}(x)&= \int \frac{\text{d}^2\vec{p}}{(2\pi)^2}\int_{-\infty}^\infty\frac{\text{d}q}{2\pi}\int_{-\infty}^\infty \frac{\text{d}P^0}{2\pi}\,
\text{e}^{\text{i} \vec{p}.\vec{r}}\text{e}^{\text{i} qz}\text{e}^{-\text{i} P^0t}\,T^{\mu\nu}(P^0,\vec{p}),\label{Eqn6.18}\\
T^{\mu\nu}(P^0,\vec{p})&=2 \pi \sum_N\delta(2p^0-m N) B^{\mu\nu}(\vec{p},N), \label{Eqn6.19}\\
B^{00}(\vec{p},N)&=\beta_N\,J^2_{N/2}(z),\quad z=\frac{|\vec{p}|}{2m},\quad\beta_N=M (-1)^{N/2}\cos\Big(\frac{N \pi}{2}\Big), \label{Eqn6.20}\\
B^{0a}(\vec{p},N)&=\beta_N\,\frac{P^0p^a}{|\vec{p}|^2}J^2_{N/2}(z), \qquad a=1,2,\label{Eqn6.21}\\
B^{ab}(\vec{p},N)&=\beta_N\Big(A_N\delta^{ab}+E_N\frac{p^ap^b}{|\vec{p}|^2}\Big), \qquad a,b=1,2,\label{Eqn6.22}\\ A_N&=\frac{1}{4}[J_{\frac{N}{2}+1}(z)\!-\!J_{\frac{N}{2}-1}(z)]^2,\quad E_N=J_{\frac{N}{2}+1}(z)J_{\frac{N}{2}-1}(z),\label{Eqn6.23}\\
B^{\mu3}(\vec{p},N)&=B^{3\mu}(\vec{p},N)\,=0,\qquad \;\; \mu=0,1,2,3.\label{Eqn6.24}
\end{align}
\indent For $e^-e^+$ pair production via a graviton emitted from the string, the amplitude of the process is given by
\begin{equation}\label{Eqn6.25}
M \propto T^{\sigma\lambda}(2p^0,\vec{p}_1+\vec{p}_2)\,\frac{[\eta_{\sigma\mu}\eta_{\lambda\nu}-\frac{1}{2}\,\eta_{\sigma\lambda}\eta_{\mu\nu}]}
{(p_1+p_2)^2} T^{\mu\nu}_{e^-e^+},
\end{equation}
\noindent where $T^{\mu\nu}_{e^-e^+}$ is the energy-momentum associated with the pair. This expression simplifies to
\begin{equation}\label{Eqn6.26}
M \propto \frac{1}{(p_1+p_2)^2}\Big\{-2\,m\,\bar{u} v T^{00}+2[(\bar{u}\gamma_av)(p_b^1-p_b^2)+ m \delta_{ab} \bar{u} v]\,T^{ab}\Big\}.
\end{equation}
\indent The recurrence relation of Bessel functions
\begin{equation}\label{Eqn6.27}
J_{\frac{N}{2}-1}(z)=\frac{2\sqrt{2}}{\beta}\,J_{\frac{N}{2}}(z)\,-\,J_{\frac{N}{2}+1}(z),
\end{equation}
\noindent allows one to express $A_N,\, E_N$ in terms of $J_{\frac{N}{2}+1}$, $J_{\frac{N}{2}}$ which differ by one order only, and for sufficiently high energies, they may be expressed in terms of $J_{\frac{N}{2}}$.\\
\indent The polarization correlations probability of the pair is then readily computed from (\ref{Eqn6.26}) to be
\begin{align}
&P[\chi_1,\chi_2]=\frac{1}{4(4-\beta^2-2\sqrt{2} \,\beta)}\times\nonumber \\
\times\,\Big\{\big(\sqrt{2(1-\beta^2)}\! -\! \sqrt{2}\! +\! \beta\big)^2&\cos^2 \Big(\frac{\chi_1-\chi_2}{2}\Big)+\big(\sqrt{2(1-\beta^2)}\! + \!\sqrt{2}\! -\! \beta\big)^2\cos^2 \Big(\frac{\chi_1+\chi_2}{2}\Big)\Big\}\label{Eqn6.28},
\end{align}

\noindent and for those involving with the measurements of only one of the spins, we again simply have
\begin{equation}\label{Eqn6.29}
P[\chi_1,-]\,=\,\frac{1}{2},\qquad P[-,\chi_2]\,=\,\frac{1}{2}.
\end{equation}

\indent For example, for $\beta=0.8$, $\chi_1=0^{\text{o}}$, $\chi_2=160^{\text{o}}$, $\chi'_1=100^{\text{o}}$, $\chi'_2=10^{\text{o}}$, as before, now give for the indicator $S=-$ 1.103, violating Bell's inequality.\\
\indent The fact that the polarization correlations probabilities of the pair $e^-,e^+$  emitted via the charged and neutral strings are different, in general, would indicate, in principle, whether the underlying string is charged or uncharged. In the extreme relativistic case $\beta \rightarrow 1$, the joint probability in (\ref{Eqn6.28}) approaches the one in (\ref{Eqn6.17}), and the joint probabilities resulting from both strings coincide.\\

\section{Conclusion}
\noindent In this paper we have reported, in detail, on \textit{explicit} quantum field theory (QFT) computations (not just in words but actual computations) of polarization correlations in several processes in QED, the Electroweak Theory and even in processes of particle productions from strings. In \textit{all} these processes, QFT shows explicit dependence on the speed (energy) of the colliding particles and this is as a rule rather than exception. After all, particles carry speeds in order to merge, collide and create other particles. One may na\"{\i}vely argue that
one may arrange such speeds to be low enough so that they become negligible. That this cannot always be done, becomes clear in our investigation of the process $e^-e^+ \rightarrow \mu^-\mu^+$ in the Electroweak Theory in Sect.5, where due to the fact that the mass of the electron is much smaller than that of the muon, the $e^-,e^+$ pair should have enough energy (threshold energy) to create the muons pair. This energy factor makes its appearance in the expression of the polarization correlation and cannot be reduced below the threshold value on physical grounds. What is equally interesting is that the polarization correlation of the muons not only depends on energy but also depends on coupling parameters ratios as well as mass ratios. On the other hand, how one can discuss locality problems, in the light of Bell's inequality, by considering low energy non-relativistic limits. Clearly investigations based simply on combining spins, or even based on non-relativistic quantum mechanical methods, become questionable. QED and the Electroweak Theory have stood the test of time, and are dynamical theories, why not use them to explicitly carry out such computations ?. After all QFT is the theory that emerges upon extending quantum theory to the relativistic regime and is based on local interactions. Needless to say, all the expressions of the relevant probabilities for polarization correlations, and of those probabilities involved in the measurement of only one of the polarizations of the emerging particles, lead to a clear violation of Bell's inequality, and support
the monumental theory we call quantum theory.\\
\indent The novel properties derived of polarization correlations, such as their dependence on energy, as explicitly predicted by QFT, obviously call for experimental tests which monitor the speeds of colliding particles. Such experiments, if realized, would, in turn, provide new tests of
these fundamental interactions devised by men.\\

\noindent \textbf{Acknowledgments}\\
 The authors would like to thank their colleagues at the Institute for their interest in this work.\\

\end{document}